# Experimental investigation of turbulent flow in a two-pass channel with different U-turn

Runzhou Liu, Haiwang Li, Ruquan You*, Zhi Tao

*National Key Laboratory of Science and Technology on Aero-Engines Aero-Thermodynamics,*

*Beihang University Beijing, 100191, China*

*Aircraft/Engine Integrated System Safety Beijing Key Laboratory,*

*Beijing 100191, China*

*Corresponding Author: (youruquan10353@buaa.edu.cn)

**Abstract:** In this paper, the TR-PIV method is used to study the internal flow field characteristics in U-shaped channels. The Reynolds number, based on the square cross section channel hydraulic diameter is 8888,13333 and 17777. Mean flow, Reynolds stress and POD are taken into consideration to investigate the flow characteristic with three different turning sections. Through analysis, a series of important conclusions have been drawn. For the main flow, the structure of turning sections has obvious influence on the characteristics of flow field. The size and number of vortices in the corner area are significantly reduced, because the increase in Reynolds number makes the influx impact stronger. It can be seen from the Reynolds stress distribution which is obviously different in different turning sections that the pulsation caused by the mixing of the main flow and the vortex is obviously stronger than that at the boundary. The flow at the turning section is complex, the distribution of the proportion of turbulent kinetic energy in the low-order mode is relatively gentle, and there is an obvious wavy structure at the turning section of the inner circle and outer circular passage, which matches the velocity field from the POD.

## I. Introduction

Turbine blade cooling is generally divided into external and internal cooling. The optimized design of the middle chord is critical to improving the cooling efficiency of the turbine blades, because it has the largest area. The researcher usually simplifies the middle chord into a model of the U-shaped channel. The study of heat transfer and flow field characteristics is also mainly focused on the U-channel turning section. From the current research, heat transfer research has been relatively mature. However, for the flow, first, the flow experiment is difficult to carry out. Second, the turbulence model can no longer meet the current calculation requirements. In order to explain the heat transfer phenomenon in the U-shaped channel and correct the turbulence model at the same time, it is convenient for subsequent calculations to facilitate the optimization design of the inner cooling channel of turbine blade. This paper will focus on the internal flow field analysis of the U-shaped channel under different structure to lay the foundation for later calculations and turbine blade design.

The research methods for U-channel flow field are the same as other flow field research methods, which are generally divided into two categories: simulation and experiment. Longfei Wang[1] et al. studied the heat transfer and flow characteristics of U-shaped channel with wavy ribs under static and rotational conditions by numerical simulation. M.R.H. Nobari[2] et al. found that in the bend region, the centrifugal forces due to curvature are significant, performing an intensive outward secondary flows resulting in an increase of heat transfer on the outer wall. But in the straight parts of the duct, secondary flows due to Coriolis forces were dominant. Guoguang Su[3] et al. used the

RANS method to study the effects of different Reynolds number and aspect ratio on flow and heat transfer in smooth rotating U-channel with small aspect ratio. In terms of experiment, in the early 21th century, many scholars[4-6] adopted the PIV method to study the flow and heat transfer in channel. YouQin Wang[7] compared the U-channel flow field structure measured by J. Martin[8] et al. using the PIV method through numerical simulation. Zhongyang Shen[9,10] et al. studied the heat transfer and friction performance of U-channels with ribs, dimples, and protrusions through numerical simulation. It was found that the heat transfer effect of the outlet section was significantly stronger than that of the inlet section. There were phenomena such as separation, reflow and secondary flow in the turning section. The heat transfer effect of the rib-protrusion structure was much stronger than that of the rib channel under rotation, especially in the inlet section. C. Brossard[11] et al. used the PIV method to study the flow field characteristics of the rotating ribbed U-channel and found that the rotation caused the heat transfer characteristics of the leading and trailing sides of the channel to be inconsistent. M. Gallo[12] et al. studied the flow and heat transfer of a smooth U-channel with a 1:1 aspect ratio. They used water as the working medium and set up an 18D (hydraulic diameter) inlet section before entering the turning section. It was found that the heat transfer was obviously enhanced at the point of fluid impact; at the same time, the heat transfer was also enhanced to a certain extent at the position where the vortex was generated. M.Gallo[13] et al. also obtained the visualizations of the main and secondary flow fields in a U-shaped channel and perform a 3D reconstruction of the mean flow and vortical fields. J.Visscher[14] et al. investigated the massively separated turbulent flows with rotation by PIV. In addition to channel, PIV was applied into open-cavity flow by J.Basley[15].

At present, the experiment on heat transfer in the U channel has also been carried out a lot in Refs[16-22]. Zhongyang Shen[17] et al. numerically studied the effect of bleed extraction on a U-shaped internal passage with dimple structures. Yang li[21] et al. experimentally investigated the effect of wall-temperature ratios(TR) and channel orientations on heat transfer in rotating smooth square U-channel. Haiwang Li[22] investigated the heat performance in a U-shaped smooth channel with engine-similar cross-section by experiment. Lei Luo[19] et al. numerically studied the effects of a dust hole and its location on heat transfer and friction in a U bend channel. Mokhtar.K[23] investigated the impact of ribs on the thermal performance of rotating U-type channel, In the current research, there are many differences in heat transfer law which cannot be explained, which is the result of different structure of U-shaped channel. The theory of the flow field is needed to explain, but it is difficult for the lack of flow data. There is little comparison in flow fields between different turning sections, so the research in this paper is needed.

Proper orthogonal decomposition (POD) has a wide range of applications in image processing, electromechanical control, and signal recognition as a method for analyzing massive data[24]. It will get a set of optimal bases called mode which occupy different proportions of turbulent kinetic energy in flow field after decomposing data through POD. This method is also commonly used to identify the structures of different energy levels in the turbulent flow field, and the structures with larger energy content are extracted for analysis. Yijia Zhao[25] et al. extracted the dynamic coherent structure using a new POD method named F-POD. F. Coletti[12,26-28] et al. studied the turbulent flow inside a channel by means of time-resolved particle image velocimetry. Sen.M[29] et al. investigated a rough-wall turbulent boundary layer in a channel by using snapshot POD. Janiga[30] et al. characterized the dominant coherent flow structures in the entire three-dimensional domain using the 3D-POD technique. Zhao wu[31] et al. investigated the post-processor analysis of direct numerical simulation of a low momentum laminar jet via POD and Dynamic Mode Decomposition (DMD). They found the POD was quite optimal in reconstructing the flow while the DMD required more modes in turn. Alvarez-Herrera.C[32] et al. studied the thermal convection in a thin two-plate channel and determined the dynamics of heat flow and energy distribution using POD. They found that when the energy was distributed among many POD modes, the fluid flow is disorganized and unsteady. In addition, POD is also used in many other fields. Banafsheh.B[33] et al. investigated the transient heat conduction in on-chip interconnects using POD method. Wu Jin[34] et al. studied the mechanism of unsteady cellular flame via POD.

Through investigation, it can be found that the current research on the U-shaped channel mainly focuses on the flow field and temperature field. But the influence of different structural turning sections on the flow field in the channel is still relatively rare. For the further analysis of the flow field, some scholars have adopted the POD, but further research and application of this method for U-channel is needed. First, this paper will build a flow field test bench, based on which the U-channel flow field of different structures will be tested. Then the POD method is used to decompose the flow field to capture the flow characteristics of each mode which has different turbulent kinetic energy.

## II. Experimental set-up
### A. Experimental facility

The test bed includes the test section, the air intake device, and the data acquisition and processing device, as shown in Fig. 1. The air intake device mainly includes air pump, cooler, flow control valve, thermal flowmeter and particle generator. The air pump draws directly from the atmosphere, and the temperature and pressure rise after the air is compressed by the air pump. The working environment set in this experiment is room temperature, so the temperature of the air needs to be lowered by the cooler. The cooler uses water as the cooling medium, and the hydraulic diameter of the gas pipeline is 50mm, including 4 flow pipelines. In order to increase the heat transfer area, a fin is added to the pipe. It is possible to ensure that the gas temperature is around room temperature by adjusting the flow rate of water and the liquid level (the temperature difference does not exceed 5 ° C).

The amount of air flow which is measured using the thermal flowmeter is controlled by a flow control valve. After the air passes through the flowmeter, the particles which are generated by a particle generator based on a Laskin-nozzle blend into the main stream and then enter the experimental part along with the air[35]. In order to ensure that the gas entering the channel is a uniform smooth flow field, it is necessary to set an inlet settle section shown in Fig. 2(a)and rectify through a multi-layer filter[36]. Fig. 2(b) shows the rectification effect of the inlet section. It represents the velocity of square-square channel at the green dotted line. The flow field still maintains uniformity near the turning section, indicating that the inlet section can ensure that the rectification can ensure the uniformity of the air flow and meet the experimental requirements.

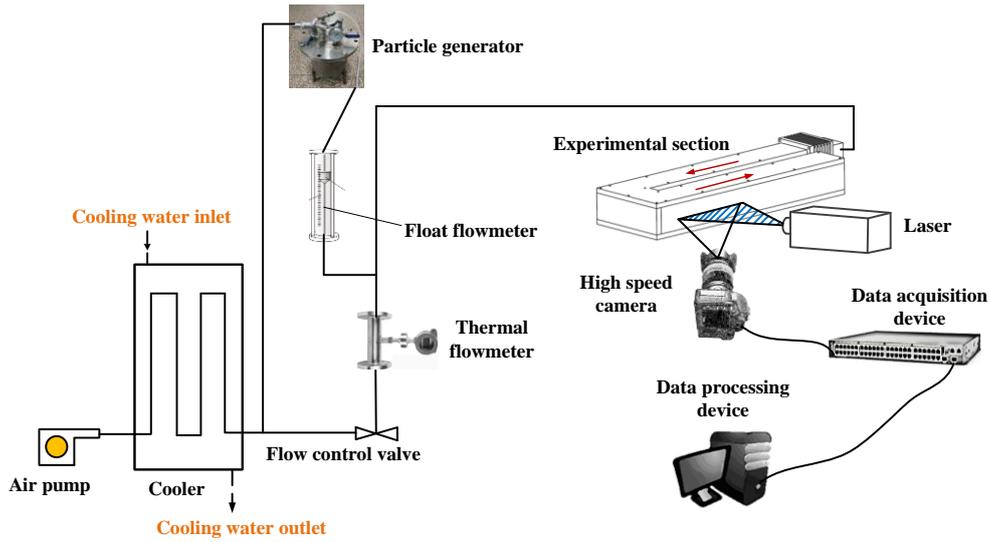

Fig. 1.Schematic diagram of test bench

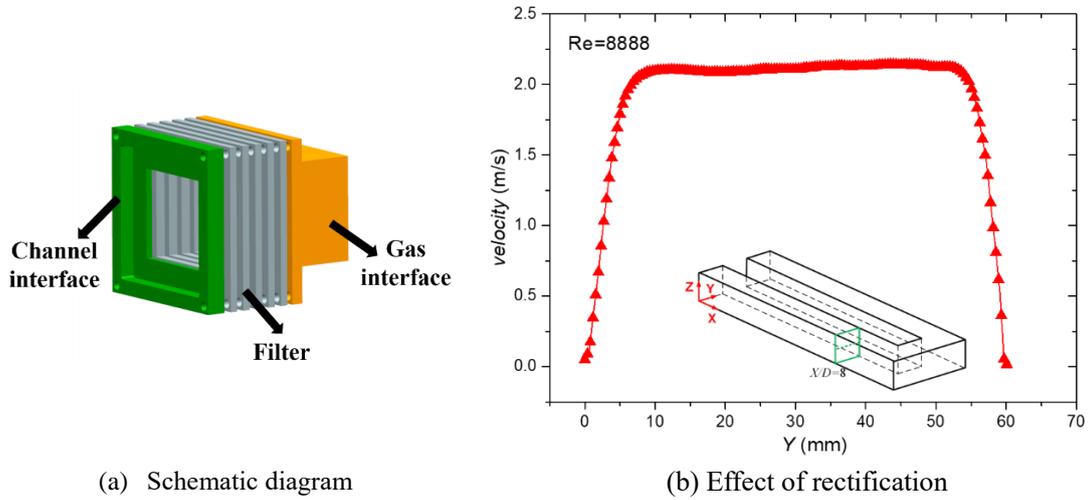

(a) Schematic diagram          (b) Effect of rectification

Fig. 2.Inlet settle section

## B. Test section

The model used in this paper is derived from the U-shaped cooling channel structure of the chord portion of the cold passage in the aero-engine turbine blade. To facilitate experimental research, we simplified the actual structure into a smooth square section U-shaped channel. The channel(600mm*180mm*60mm) shown in Fig. 3 is enlarged for improving the resolution of the flow measurement. We define $\alpha = X/D$ and $\beta = Y/D$. X and Y are the horizontal and vertical coordinates of the position of the plane, and D(60mm) is the length of the cross section of the channel. The channel can be divided into three parts: inlet section($0 < \alpha < 10$, $0 < \beta < 1$)、turning section($9 < \alpha < 10$, $1 < \beta < 2$) and outlet section($0 < \alpha < 10$, $2 < \beta < 3$). The high-speed camera is fixed under the channel and the laser is located on the side of the channel. The laser illuminates the middle of the flow field ($Z/D = 0.5$) and adjusts the camera lens to focus on this area. When the flowmeter display is stable, the data acquisition device can be turned on for acquisition, and then the flow field data can be obtained through the data processing device.

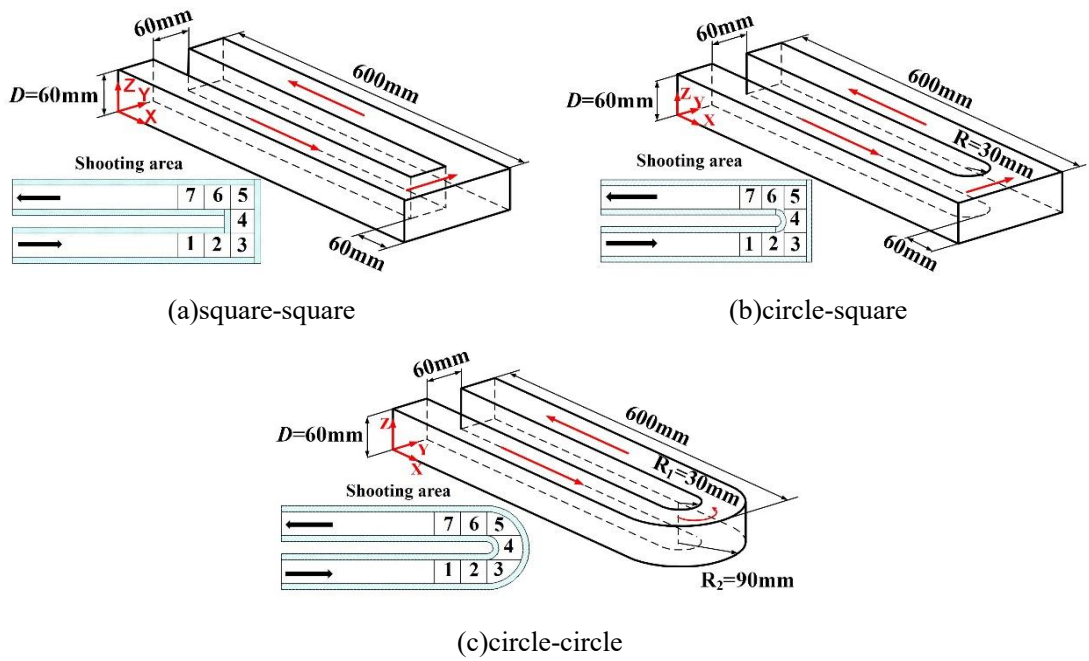

(a) square-square  (b) circle-square

(c) circle-circle

Fig. 3. Schematic diagram and shooting area of U channel

**C. TR-PIV setup**

With a high-speed camera, it is possible to take two photos of laser-illuminated particles with very short time intervals and obtain the displacement of the same particle on two frames of photos through the algorithm. Since the shooting rate can be artificially set, the velocity field can be calculated from the displacement. In the current work, the laser provides a continuous 10W 532nm sheet light which has a thickness of less than 1mm. table 1 lists the experimental parameters at different conditions.

| Parameters | value |
| --- | --- |
| Measuring position | $7<X/D<10$, $0<Y/D<1$, $Z/D=0.5$ |
| Re | 8888, 13333, 17777 |
| Frames per second | 6400, 6400, 8100 |
| Delay time ($\Delta t$) (μs) | 156, 156, 123 |
| Frames | 10000 |
| Magnification factor (mm/pixel) | 0.0787353 |

**D. Introduction of POD**

Proper orthogonal decomposition (POD) is used in this experiment to identify the spatial characteristics of the superimposed flow fields. The mathematical idea of POD can be summarized as, for a group of known functions, we try to find a group of basic functions called mode closest to it in the mean sense. A detailed introduction about POD is summarized in Refs[37-39]. In the present work, since the flow data is limited, a quick overview of POD is more suitable as compared with the classical approach. Sirovich[40] aimed at the problem that the spatial correlation matrix is difficult to be solved directly when the number of spatial points is large in the direct method. In 1987, he proposed the snapshot POD method, a new mathematical processing method, which is generally applicable to the case that the number of spatial points *m* is much larger than the sampling number *N*. Assuming that the number of sampling points in space is m, N sets of spatial data samples can be obtained after sampling N times: $(u_1(x_1), u_1(x_2), \cdots, u_1(x_m))$, $(u_2(x_1), u_2(x_2), \cdots, u_2(x_m)), \ldots, (u_N(x_1), u_N(x_2), \cdots, u_N(x_m))$. It is translated as matrix:

$$U = \begin{bmatrix} u_1(x_1) & u_2(x_1) & \cdots & u_N(x_1) \\ u_1(x_2) & u_2(x_2) & \cdots & u_N(x_2) \\ \vdots & \vdots & \vdots & \vdots \\ u_1(x_m) & u_2(x_m) & \cdots & u_N(x_m) \end{bmatrix}$$

The snapshot method subtly transforms an *m*-order high-dimensional matrix into an *N*-order low-dimensional matrix to make the calculation greatly simplified. The correlation matrix of U is defined as: $C = \frac{1}{n} U^T U$. The matrix *C* describes the time dependence of the flow field at two moments $((u_i(x_1), u_i(x_2), \cdots, u_i(x_m)), (u_j(x_1), u_j(x_2), \cdots, u_j(x_m)))$. Solving the eigenvalues of the time correlation matrix C: $CA = \lambda A$, we can get a set of eigenvalues $(\lambda_1 \geq \lambda_2 \geq \cdots \geq \lambda_N \geq 0)$ and corresponding eigenvectors $(A = (A_1, A_2, \cdots, A_N))$. Constructing a characteristic function: $\Phi = U \cdot A$, dividing each column by $\sqrt{\lambda_i} (i = 1, 2, \cdots, N)$, then get a set of orthonormal basis $\{\varphi_1, \varphi_2, \cdots, \varphi_N\}$. This orthonormal basis is the eigen function, which is the mode we analyze[41,42].

### III. Discussion of uncertainties

The uncertainty of the velocity term is the uncertainty of the flow velocity measured by TR-PIV under nonrotation. The velocity measured by TR-PIV is: $V_{abs} = \frac{M \Delta s}{\Delta t}$. Where $\Delta t$ is the time interval between two frames, which is the artificially set camera shooting speed, the error can be ignored; M is the magnification factor (pixel/mm), and the error can be neglected; $\Delta s$ is The displacement of the particles is calculated by the TR-PIV system. So, the error is mainly from $\Delta s$ in this experiment. The particle size used in this experiment is 1 μm. If the shooting speed is too low, particle smearing will occur, resulting in calculation errors. Considering the above factors, it is necessary to carefully consider the shooting speed, and the particles occupying 8 to 10 pixels per frame are optimal. The shooting accuracy of this experiment can be considered as 0.1 pixel[43]. The velocity measured by PIV is the true velocity of particles in the channel, so:

$$\Delta V = \Delta V_{abs}$$

$$\Delta V_{abs} = \sqrt{\left(\frac{\partial V_{abs}}{\partial M}\right)^2 \cdot (\Delta M)^2 + \left(\frac{\partial V_{abs}}{\partial (\Delta s)}\right)^2 \cdot (\Delta(\Delta s))^2 + \left(\frac{\partial V_{abs}}{\partial (\Delta t)}\right)^2 \cdot (\Delta(\Delta t))^2}$$

From the above analysis, the error of M and $\Delta t$ can be neglected; and $\Delta s$ can be considered as

0.1 pixel. Therefore, the speed uncertainty of this experiment is: $\Delta V = \Delta V_{abs} = \dfrac{M}{\Delta t} \times 0.1$. According to Table 1, the experimental error of the mainstream shooting of this experiment is:

Table 2 Uncertainty of mainstream

| Re | 8888 | 13333 | 17777 |
|---|---|---|---|
| Uncertainty $\Delta V$ (m/s) | 0.050 | 0.064 | 0.064 |

## IV. Result and discussion

This part focuses on the velocity and Reynolds stress field in the U-shaped channel under different turns and different Reynolds numbers. At the same time, in order to understand the detailed characteristics of the flow field, we perform POD on the flow field and analyze the first-order modes.

### A. Mean flow statistics

There are three kinds of channels with different structures, and the experimental variable of each channel is the Reynolds number. The Reynolds number can be calculated as follows: $Re = \rho U d / \mu$, where $U$、$\rho$、$\mu$ is the velocity, density and viscosity coefficient of the fluid respectively, and d is the characteristic scale. The temperature, pressure, and composition of the air can be considered constant, so the different Reynolds number represents the change in air flow rate. The size of the U channel is constant, so the different Reynolds number also represents the change in air mass flow. To better understand the flow field in U channel, it is necessary to give the mean flow field. Next, we will focus on analyzing the mean flow contours

Considering the shooting area of the camera, we divided the shooting into seven areas in total. The flow characteristic of each channel with different Re is roughly similar, so we only select one situation (square-square channel: Re=8888, circle-square channel: Re=13333, circle-circle channel: Re=17777) for each channel to analysis.

The contour for mean velocity at Re=8888 in the square-square channel is shown in Fig. 4. The air flows in from the inlet, and the flow field is uniform and smooth because of the settle section. After a period, when entering the turning section, the outer wall surface is right-angled, resulting in reflux. Therefore, there are obvious vortexes in the two corner areas. Different from area 4, in addition to the largest vortex in corner (vortex No. 1), there are two other different vortices in size in area 3, which are named as the No. 2 vortex (located in $\alpha = 9.25$) and the No. 3 vortex(located in $\alpha = 9.8$). The reflux of No.1 vortex encounters the upstream gas, then No.1 vortex turns to the downstream, and the reflux of upstream gas forms No.2 vortex. As shown in Fig. 4, the flow in the downstream direction of No.2 vortex meets the backflow of No.1 vortex, which causes the No.3 vortex. When airflow turns and passes through area 4, airflow will be far away from the inner wall due to centrifugal force, resulting in a low-pressure zone at inner wall in area 4, which leads to downstream airflow backflow. This is also the reason why backflow exists at the inner wall in outlet section. As can be seen from the contour, the recirculation zone of outlet section has been extended to area 7. The contour for mean velocity at Re=13333 in the circle-square channel is shown in Fig. 5. The reason for the recirculation zone in area 3 is like the square-square channel and will not be described here. When the airflow passes through the area 4, the airflow will be away from the inner wall surface due to the centrifugal force, thereby forming a low-pressure zone in the inner wall surface of this area, causing the downstream airflow to flow back. Since the inner wall of this channel has a circular structure, the backflow is hindered, so the recirculation zone exists only in the downstream. At the same time, the circular inner wall surface structure of the turning section causes the pressure reducing area at this point to be smaller than the square-square channel, the upstream and downstream pressure difference is reduced, and the reflow tendency is weakened, so only area 5 and 6 have a recirculation zone. The contour for mean velocity at Re=17777 in the circle-circle channel is shown in Fig. 6. Although this channel has no corner area, there is still a backflow phenomenon in the area 3 near the outer wall surface, and no reflow occurs in the area 5. The reason for the recirculation zone in the area 4 is like the circle-square channel and will not be described here.

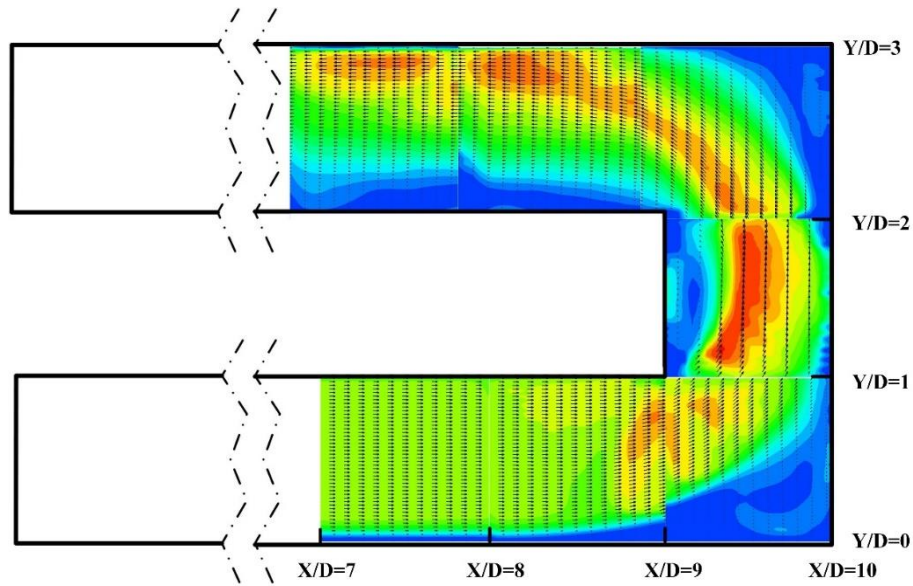

Fig. 4.Contour for the mean velocity at Re=8888 in square-square channel

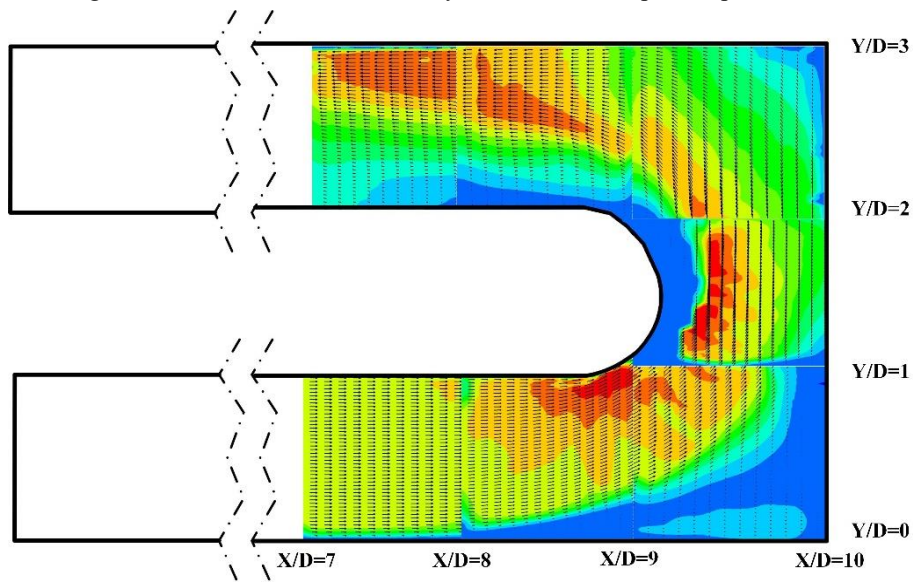

Fig. 5.Contour for the mean velocity at Re=13333 in circle-square channel

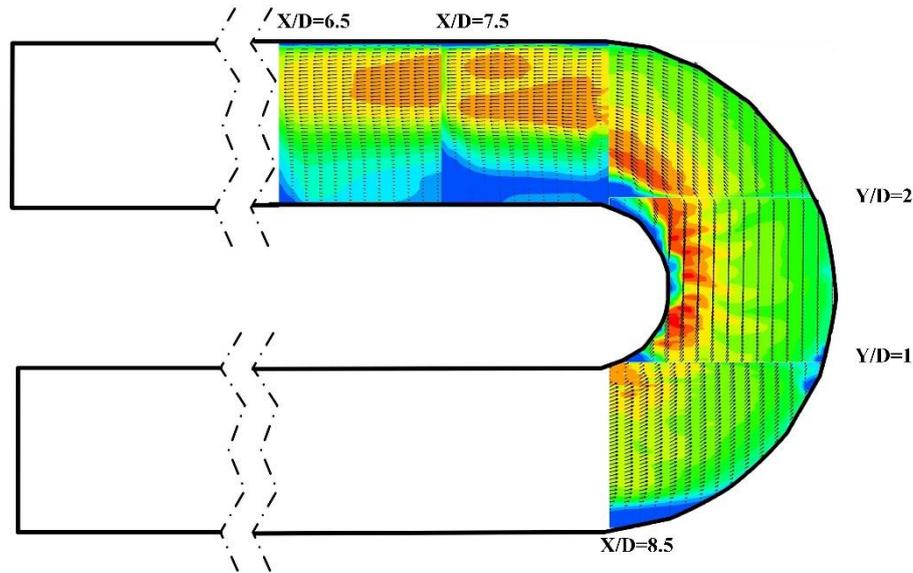

Fig. 6. Contour for the mean velocity at Re=17777 in circle-circle channel

From Fig. 4 to Fig. 6, the flow field in the turning section of the three structures is quite different. The circular structure is more consistent with the curved flow field caused by centrifugal force, and it is less likely to generate vortex in the corner and inner wall. In order to better compare the influence of different structural turning sections on the flow field, the velocity profile of some specific sections will be analyzed. We need to extract the speed from the contour and then perform dimensionless. Before that, we need to calculate the mainstream average speed under different Reynolds numbers. The specific parameters can be seen in Table 3. Fig. 7 shows and discusses the velocity profile characteristics for different Reynolds numbers and different channel. The interception position(dotted red line) is the inlet section $\alpha = 9$ ($\alpha = 9$, $0 < \beta < 1$), the turning section $\beta = 1.5$ ($9 < \alpha < 10$, $\beta = 1.5$), and the exit section $\alpha = 9$ ($\alpha = 9, 2 < \beta < 3$).

It can be seen from Fig. 7 that the velocity profile (c)(f)(i) intercepted at the turning section of the three channels with different Reynolds numbers has a high similarity. There is a peak in the velocity profile at the inner wall surface, indicating that there is one vortex here in Fig. 7(c). Fig. 7 (a)(d)(g) gives the velocity profile at the exit section of the three channels. In general, the streamlines of the three structures here are roughly similar, except that the position of vortexes on the inner wall of different structures leads to different flows on the near wall. Fig. 7(b)(e)(h) shows the velocity profile at the inlet section of the three channels. Although the Reynolds number is different, the velocity pattern has a high similarity. The reason for the existence of a maximum value and a minimum value near the outer wall surface in the case of Re=8888 and 13333 (Fig. 7(e))is that there exists recirculation zone. When the Reynolds number is increased to 17777, there is no recirculation zone, so there are no minimum values and maximum values.

Table 3 Test information under different Reynolds numbers (1atm, 20 ° C)

| Re | Dynamic viscosity (Pa·s) | Density (kg/m³) | Hydraulic diameter (m) | Velocity (m/s) |
|---|---|---|---|---|
| 8888 | $17.9 \times 10^{-6}$ | 1.1691 | 0.06 | 2.27 |
| 13333 | $17.9 \times 10^{-6}$ | 1.1691 | 0.06 | 3.40 |
| 17777 | $17.9 \times 10^{-6}$ | 1.1691 | 0.06 | 4.54 |

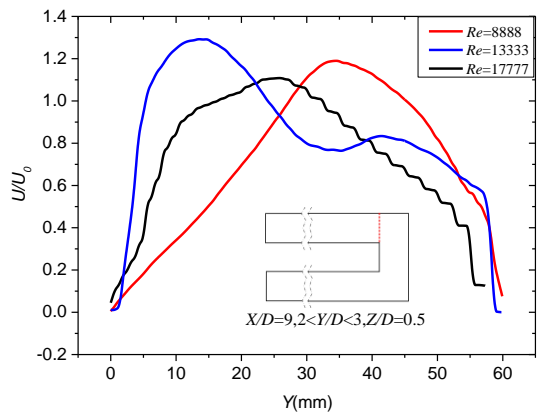

(a)

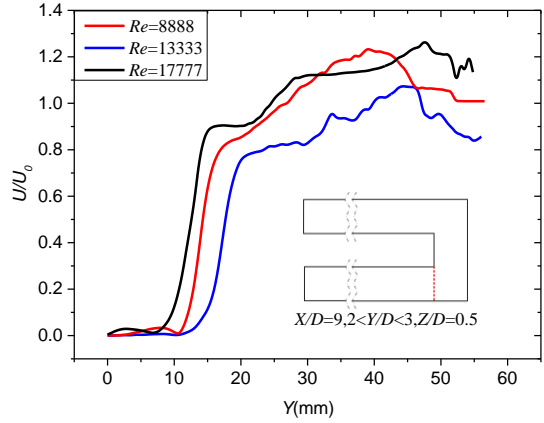

(b)

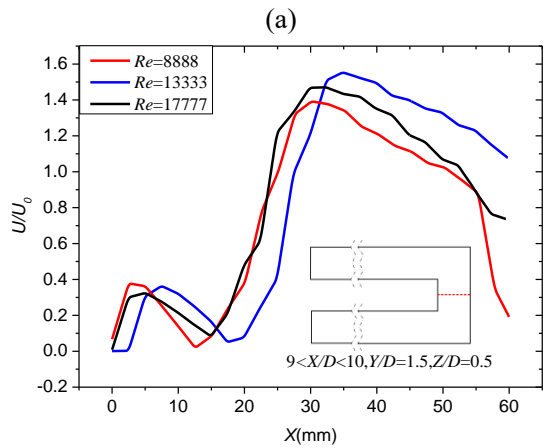

(c)

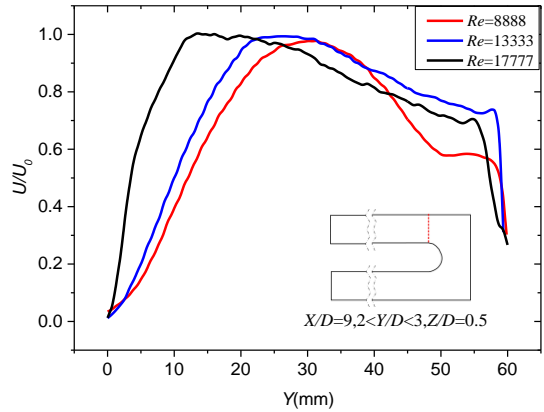

(d)

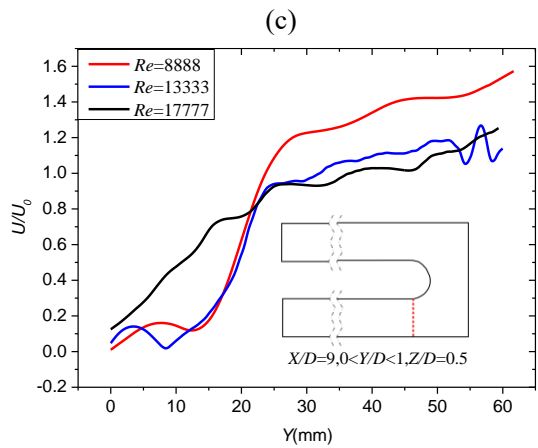

(e)

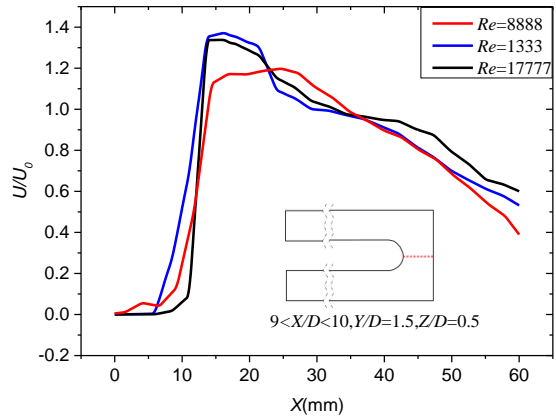

(f)

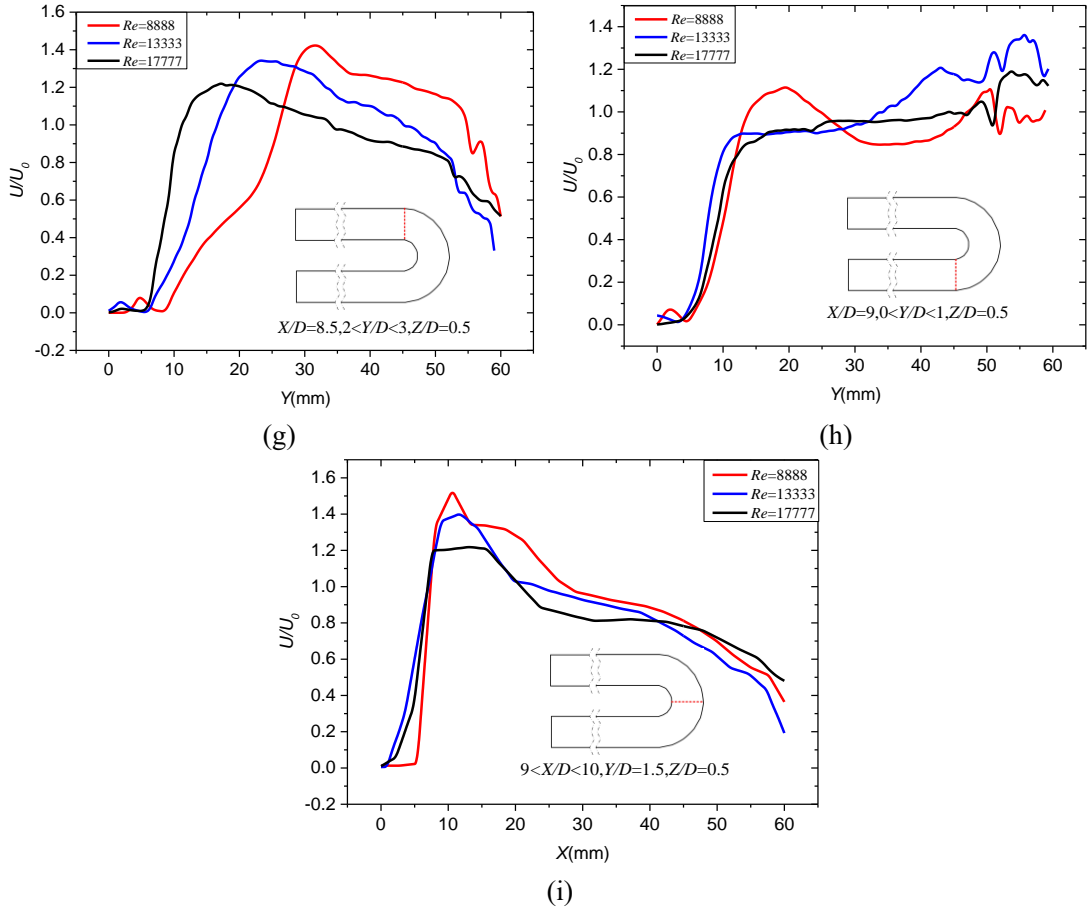

Fig. 7. Velocity profile ((a) to (c) is the square-square channel, (d) to (f) is the circle-square channel, and (g) to (i) is the circle-circle channel)

**B. Reynolds stress analysis.**

The turning section makes the distribution of Reynolds stress ($\overline{U'V'}$) more complicated. Fig. 8 shows the Reynolds stress profile of the area 4 ($9 < \alpha < 10, \beta = 1.5$) of each channel. Fig. 8(a) shows the Reynolds stress distribution of the square-square channel. The extreme value of the Reynolds stress exists in the middle position of this area, because the flow here is more complicated, the turbulence is higher, and there is a vortex structure on the inner wall surface, so the turbulence Pulsation is enhanced. The reason why there is a maximum value of the Reynolds stress near the inner wall surface of the circle-square channel is that the main stream is close to the wall surface, and the disturbance is increased, which is consistent with the cause of the other maximum value near the outer wall surface. For the circle-circle channel, the Reynolds stress is positive and there are three maximum values. It can be seen from Fig. 9 that the Reynolds stress distribution at the turning section of this channel is special, and the Reynolds stress in the upstream region near the inner wall surface is positive, and there is a stripe turbulent structure; the downstream regions are all negative, and similar stripe structure. In the half channel region close to the outer wall surface, the overall Reynolds stress is positive, and there is no stripe turbulent structure.

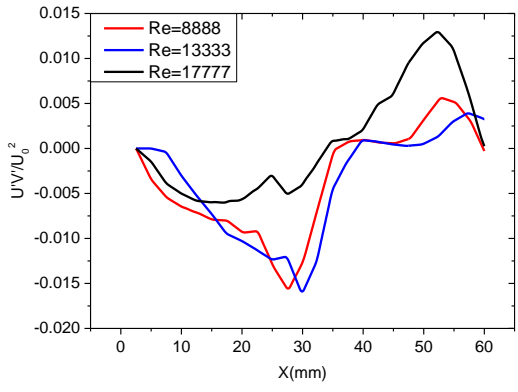
(a) square-square channel

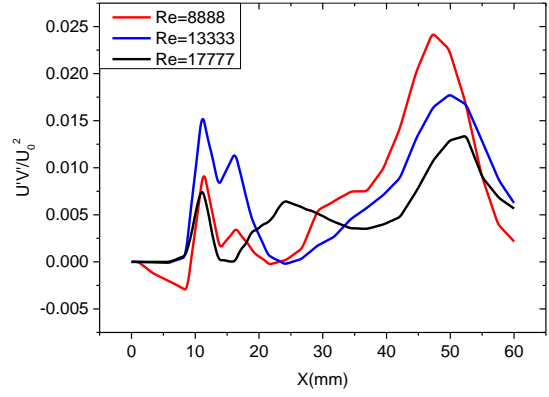
(b) circle-square channel

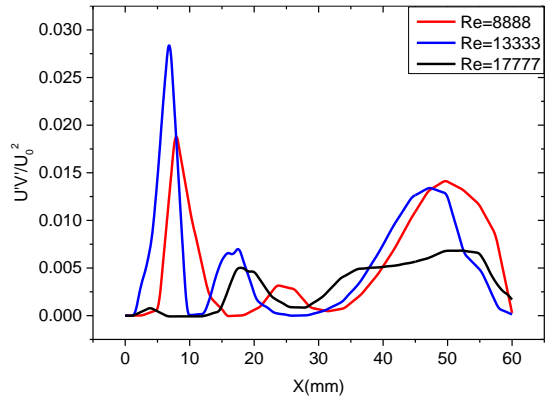
(c) circle-circle channel

Fig. 8. Reynolds stress profile of the turning section (Area 4, $9 < \alpha < 10, \beta = 1.5$)

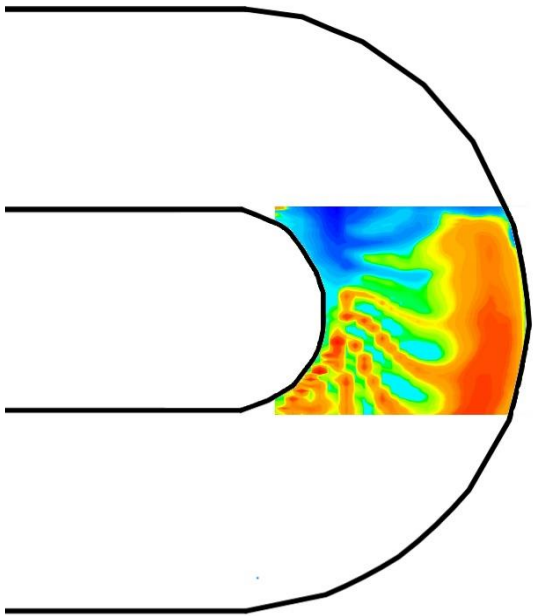
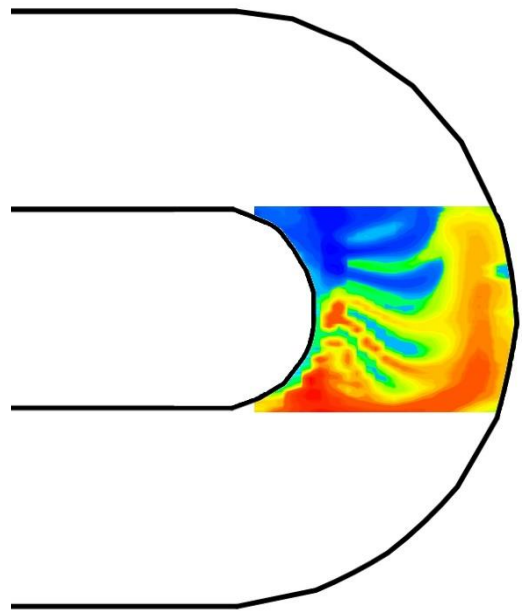

(a) Re=8888　　　　　　　　　　　　　(b) Re=13333

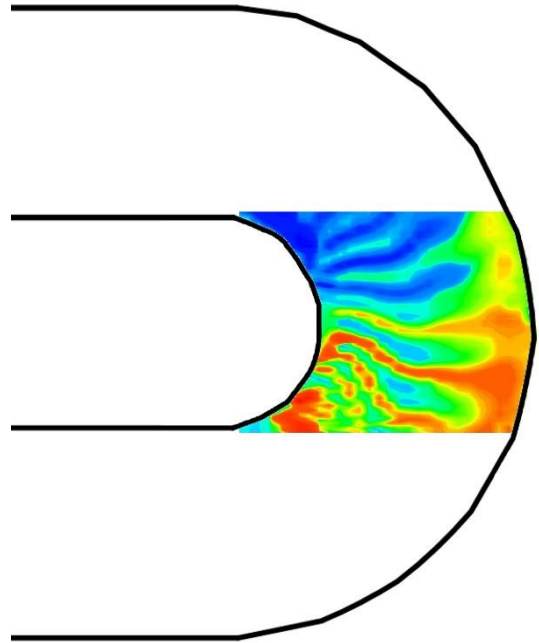

(c)Re=17777

Fig. 9.Reynolds stress contour of circle-circle channel in turning section (Area 4)

**C. Proper orthogonal decomposition**

We decompose wall-normal velocity component here, as it reflects micro flow features of various modes more clearly[26]. By capturing these features, it is possible to obtain flow characteristics of various mode that account for different flow kinetic energy ratios. The low-order mode captures most of the energy of the flow field, which also determines the basic state of the flow. High-order modes capture less energy and are small-scale coherent structures. In the real flow field, the more complex the flow distribution, the more modes are needed to capture a certain amount of energy. Considering that the flow field of the turning section is the most complicated in the U channel, POD analysis is performed on this section (area3,4,5). This paper presents the POD analysis contour of wall-normal velocity component with streamline distribution to better reveal the flow distribution of each mode. Since each mode contains different energies, the legend for the analysis results will be adjusted for different situations. This analysis selected different Reynolds numbers for different channel: Re=8888 square-square channel, Re=13333 circle-square channel and Re=17777 circle-circle channel.

Fig. 10 to Fig. 12 show the modes of the square-square channel in the 3, 4, and 5 area of Re=8888, where (a) to (d) are the first, fifth, fifteenth, and thirtieth modes respectively, (e) is the average of wall-normal velocity field. In order to more clearly compare the different Reynolds stresses of different structures, we select more representative area 3, 4, and 5 and give contour maps. Different from the area 3, the velocity direction of 1st mode in the area 4 is roughly opposite to the mean flow. The fifth mode has two symmetrical vortex structures ($\alpha = 9.4$), and in the 15 and 30 modes, smaller scale vortexes appear. The velocity core of the 1st mode in area 5 is further downstream than the mean flow. the backflow on the inner wall surface is more obvious. In the mean flow, there is only one large-scale vortex in the corner. In the contrast, there are many small-scale vortices in the 5th mode. The main reason is that the turbulent energy dissipated, and the large scale gradually breaks into small scale. For the 15th and 30th modes, the number of small-scale vortices is more, and not only in the corner. Table 3 shows the number of modes and ratio mode numbers of the three area in the square-square channel at Re=8888 when capturing 95% of energy. Approximately the first 36% mode can basically characterize the information of the flow field.

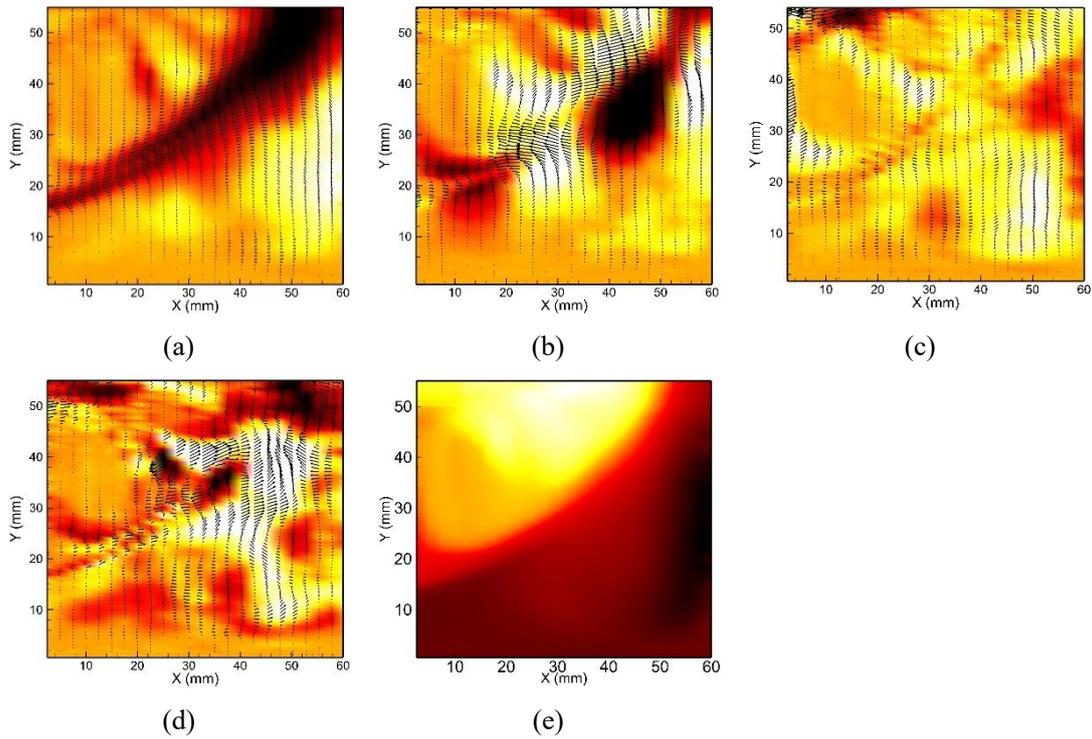

Fig. 10.First, fifth and fifteenth POD modes of the wall-normal velocity component (top row); thirtieth mode and mean of the wall-normal velocity component (bottom row). (Area 3 of square-square channel)

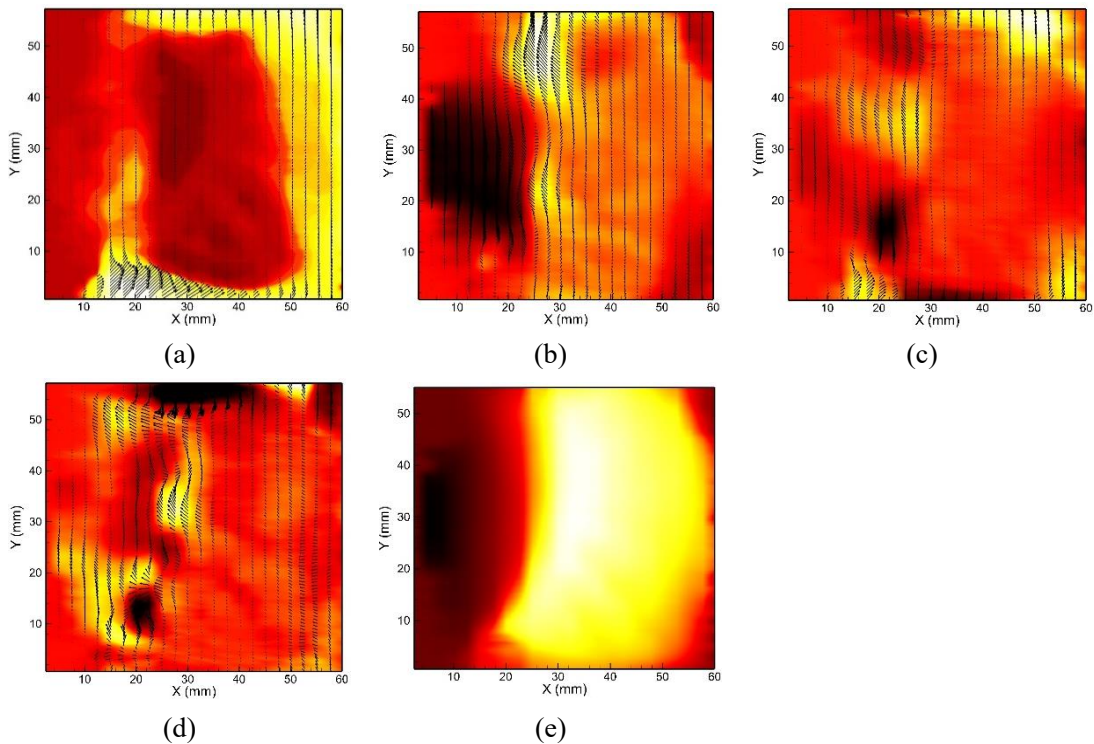

Fig. 11.First, fifth and fifteenth POD modes of the wall-normal velocity component (top row); thirtieth mode and mean of the wall-normal velocity component(bottom row). (Area 4 of square-square channel)

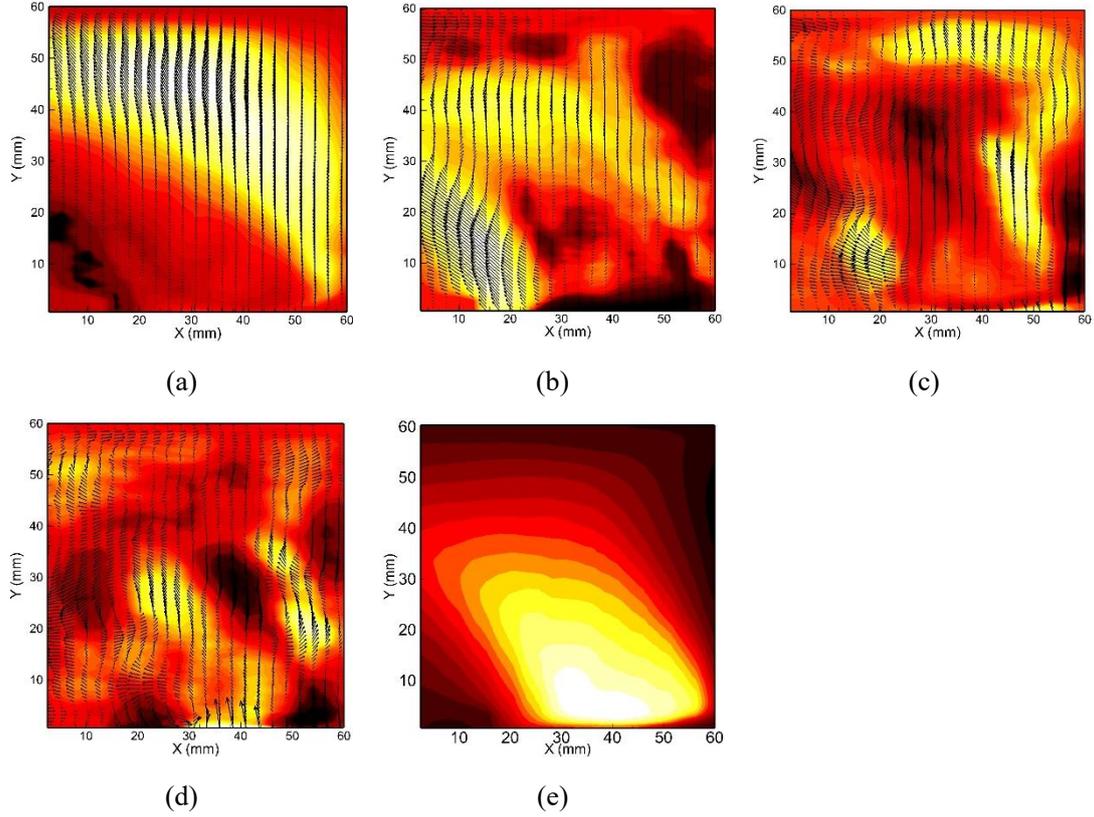

(a) (b) (c)

(d) (e)

Fig. 12.First, fifth and fifteenth POD modes of the wall-normal velocity component (top row); thirtieth mode and mean of the wall-normal velocity component (bottom row). (Area 5 of square-square channel)

Table 3 Mode information of square-square channel

| Position | The number of modes | Number of modes when capturing 95% of energy | Ratio of mode numbers when capturing 95% of energy |
| --- | --- | --- | --- |
| Area 3 | 1713 | 704 | 41.09 |
| Area 4 | 1825 | 623 | 34.14 |
| Area 5 | 2216 | 751 | 33.89 |

Fig. 13 to Fig. 15 show some modes in area 3, 4, and 5 of the circle-square channels at Re = 13333. For the first mode, except for the presence of a vortex in the corner, the other phenomena are significantly different from the average flow field. For the second mode, the corner vortex is not obvious. On the contrary, there are small-scale vortexes structure on the boundary between the main flow field and the corner vortex. For the 5th and 10th modes, the vortex structure distribution is more dispersed and the scale is smaller. For the area 4, the flow at the turning section is more complicated than that of the area 3 region. The obvious difference between the low-order mode and the mean flow is the result of that the turbulent energy is not concentrated in the low-order mode. The flow field near the inner wall of the 1st mode is like stripe. In the contrast, the whole flow field near the outer wall is not. In the high-order mode, the flow field has a low ratio of turbulent kinetic energy, and its stripe structure is not obvious. The inner wall surface of this channel has a circular structure which blocks the backflow of the downstream flow, so there is no vortex. According to the mean flow field in the area 5, only one vortex is generated in the corner. However, it can be seen from the POD decomposition results that there are a lot of small-scale vortexes, which do not only exist in the corner area. Table 4 shows the number of modes and ratio mode numbers of the three area in the circle-square channel at Re=13333 when capturing 95% of energy. Approximately the first 43% mode can basically characterize the information of the flow field.

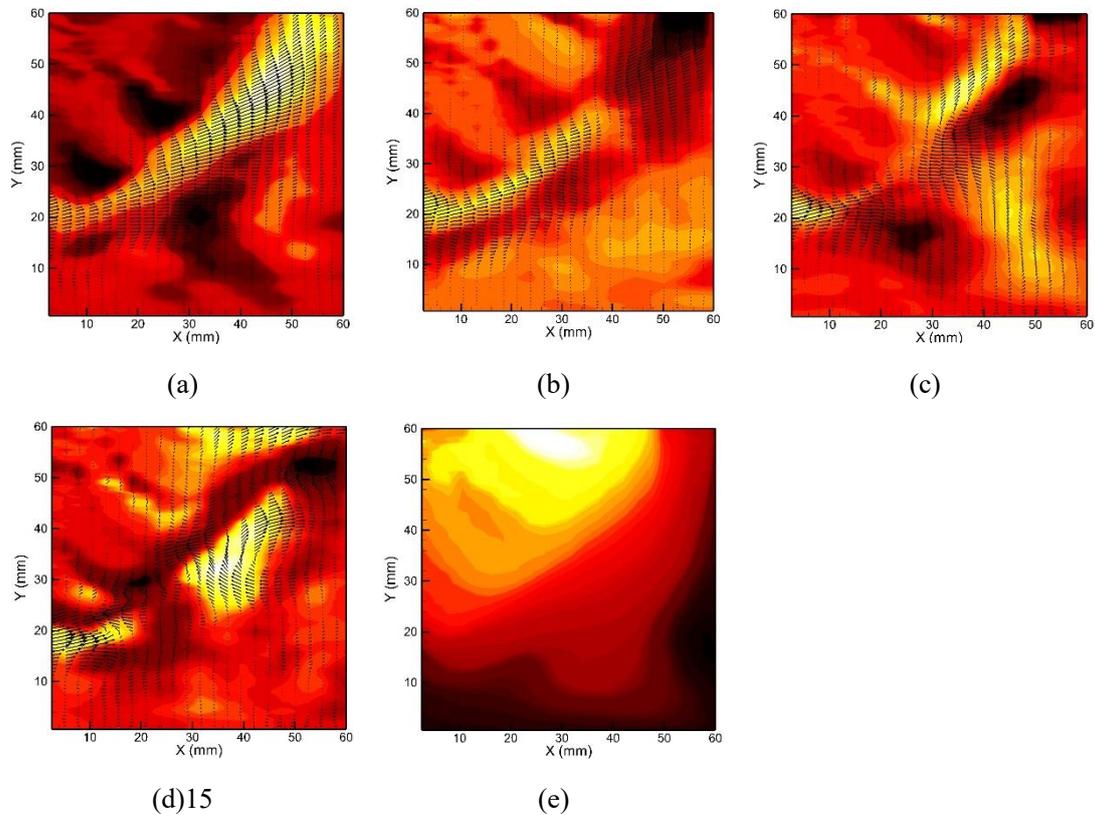

Fig. 13.First, second and fifth POD modes of the wall-normal velocity component (top row); tenth mode and mean of the wall-normal velocity component (bottom row). (Area 3 of circle-square channel)

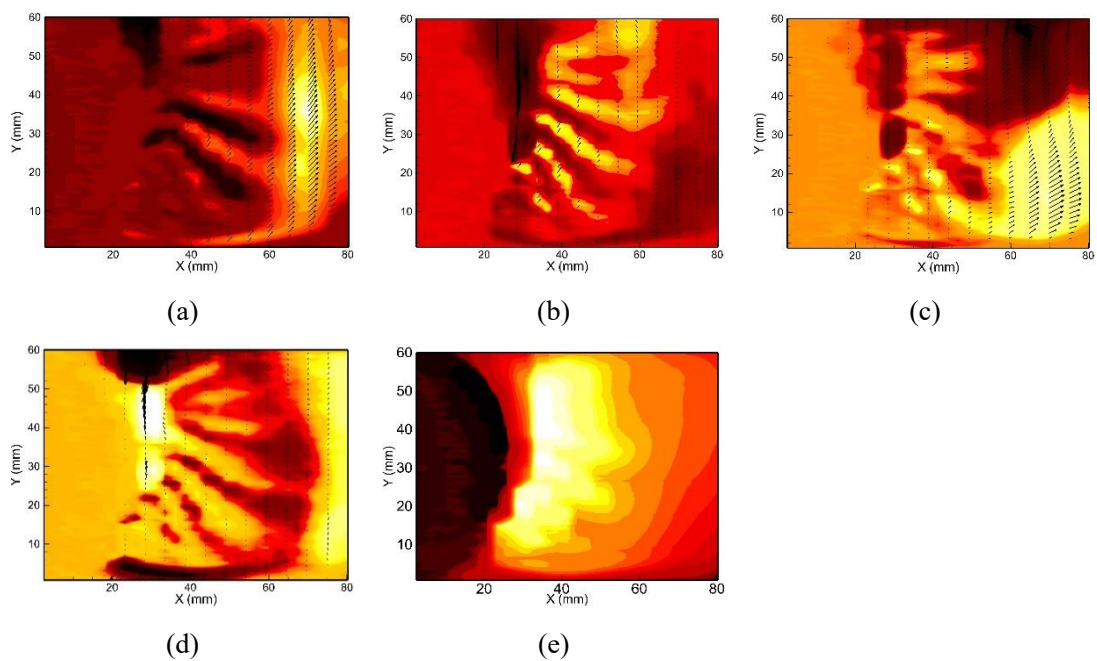

Fig. 14.First, second and third POD modes of the wall-normal velocity component (top row); fourth mode and mean of the wall-normal velocity component (bottom row). (Area 4 of circle-square channel)

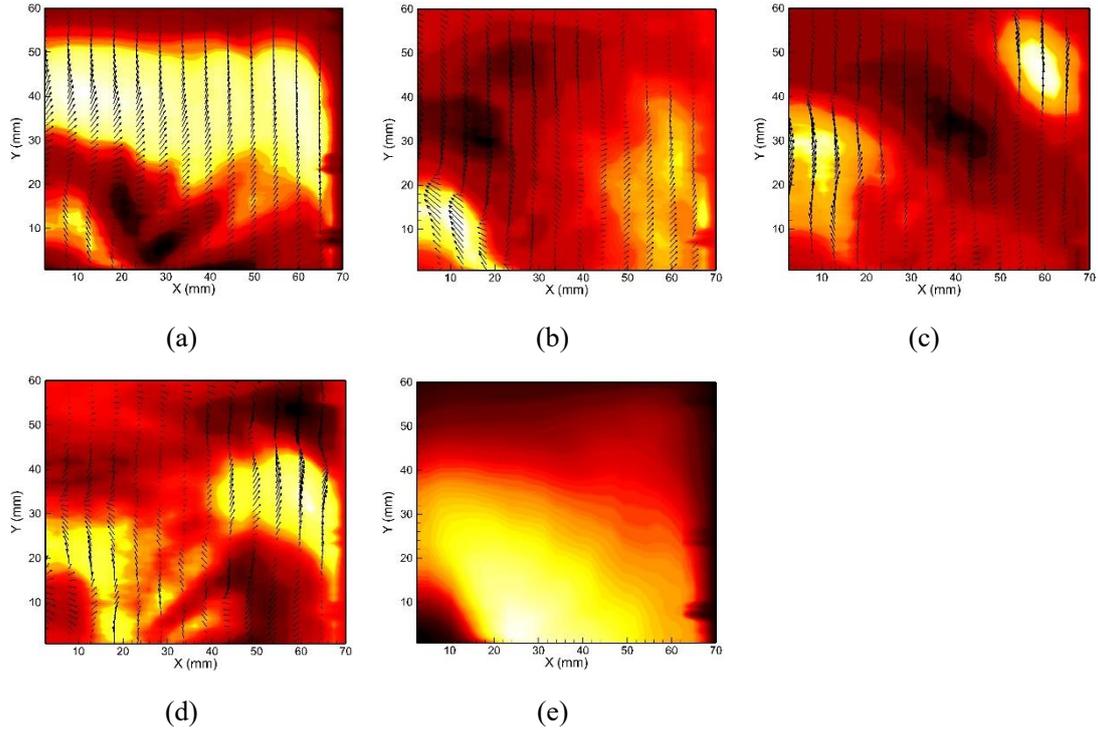

(a)           (b)           (c)

(d)           (e)

Fig. 15.First, second and fifth POD modes of the wall-normal velocity component (top row); tenth mode and mean of the wall-normal velocity component (bottom row). (Area 5 of circle-square channel)

Table 4 Mode information of circle-square channel

| Position | The number of modes | Number of modes when capturing 95% of energy | Ratio of mode numbers when capturing 95% of energy |
| --- | --- | --- | --- |
| Area 3 | 2102 | 938 | 44.6 |
| Area 4 | 1964 | 832 | 42.4 |
| Area 5 | 2207 | 923 | 41.8 |

Fig. 16 to Fig. 18 show some modes in area 3, 4, and 5 of the circle-circle channels at Re = 17777. As can be seen from Fig. 16, the result of the 1st mode in circle-circle channel is quite different from the mean flow. The first mode has a distinct velocity boundary near the wall, and the velocity signs on opposite sides of the interface are opposite and the values are approximately equal. This phenomenon does not exist in the subsequent modes, but the vortex structure gradually appears, and two vortices are apparently present in the fifth mode. A similar stripe structure exists in the circle-square channel of area 4. In high order modes, turbulent kinetic energy gradually dissipates, and stripe structure is no longer obvious. Fig. 18 shows the POD analysis results of the modes 1, 2, 3, 4, 5, 10 and 15 in area 5. The first mode has a velocity boundary at the lower left corner, and the speed signs on both sides are opposite. There is a large area of negative velocity region outside the interface, which is exactly the opposite of the mean flow field distribution. For the 3rd and 5th modes, there is a velocity boundary like the area 3, and the velocity distributions on both sides of the interface are opposite in sign and equal in value. Different from the area 3, the area 5 have interface in the 3rd and 5th modes, and the area 3 have only appeared in the 1st mode. For the 10th and 15th modes, the velocity boundary still exists, but the large-scale vortex structure on both sides of the interface gradually breaks into small-scale. It is assumed that this velocity boundary is due to the secondary flow.

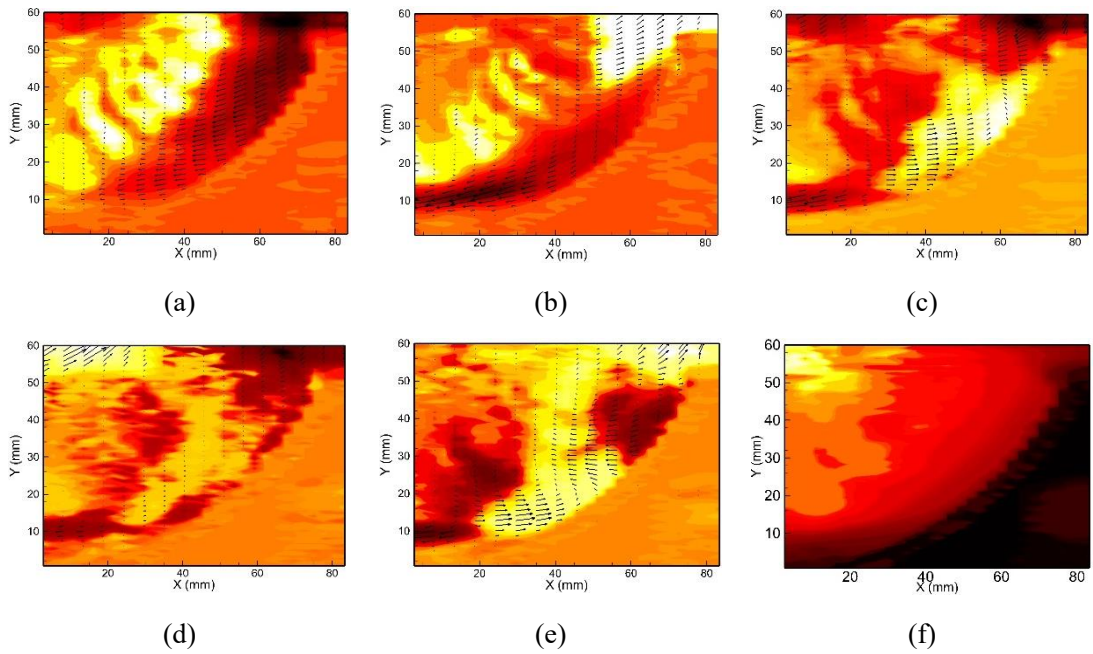

Fig. 16. First, second and third POD modes of the wall-normal velocity component (top row); fourth, fifth mode and mean of the wall-normal velocity component (bottom row). (Area 3 of circle-circle channel)

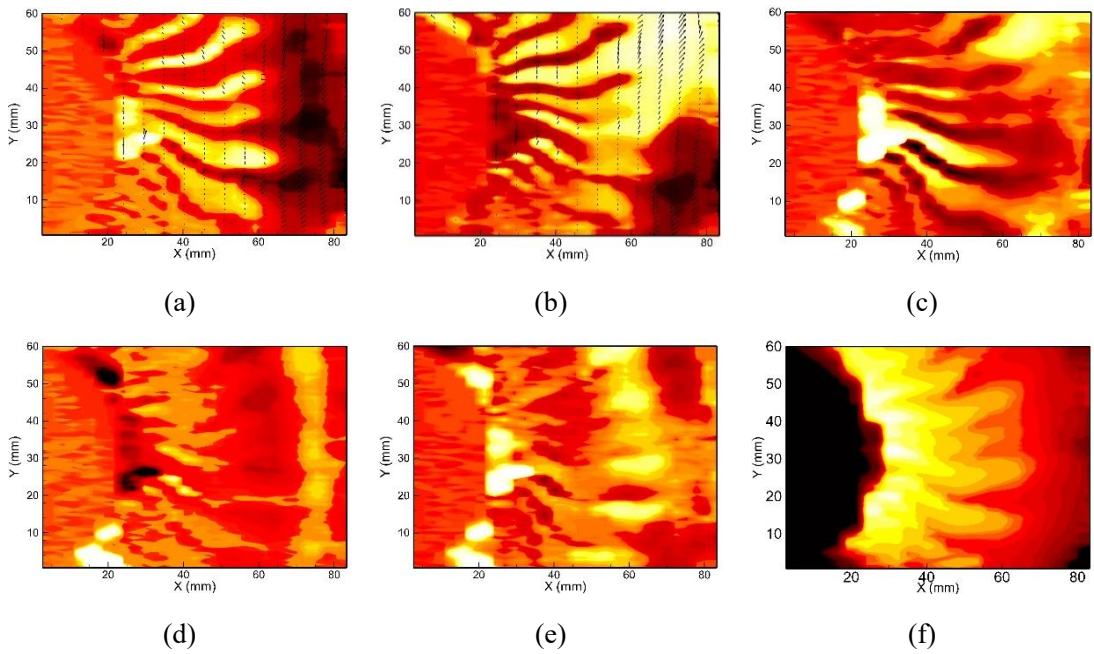

Fig. 17. First, second and third POD modes of the wall-normal velocity component (top row); fourth, fifth mode and mean of the wall-normal velocity component (bottom row). (Area 4 of circle-circle channel)

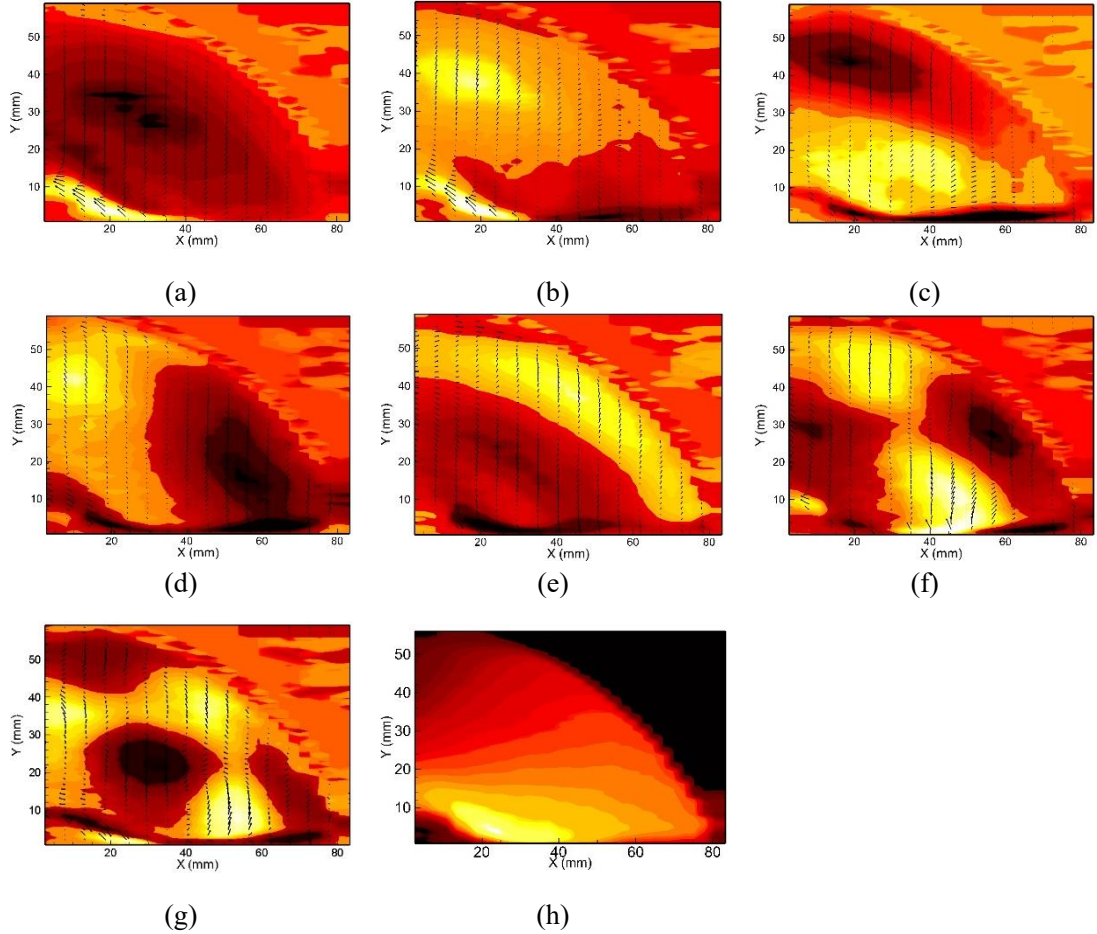

Fig. 18.First, second and third POD modes of the wall-normal velocity component (top row); fourth, fifth and tenth POD modes (middle row); fifteenth mode and mean. (Area 5 of circle-circle channel)

## V. Conclusion

This experiment reports the TR-PIV data of the stationary U-channel. Considering different Reynolds numbers, U-channels with different structures are used, and the POD method is performed to identify the spatial characteristics of the flow field. Based on the results, the following conclusions are drawn.

(1) The number, position and strength of vortexes are different with different structure. The impact of inlet flow is stronger at higher Re and the core of reflux zone flows downstream. The circular wall structure can reduce the corner vortex at the turning section, but the energy distribution reveals that the flow of the circular wall structure is more complicated.

(2) The Reynolds stress increases obviously in the mixture of reflux and main flow. The circular wall will create a stripe flow structure. The first mode occupies the most energy and is most like the average flow field. Under high order modes, large scale vortex structures dissipate into small scale structures.

(3) There is a velocity dividing line in the area 5 of circle-circle channel. The velocity distributions on both sides of the interface are opposite in sign and equal in value. The flow in circle-circle channel is more complex, so the causes of this phenomena may be related to the secondary flow in the turning section.

**Acknowledgements**

The present work is financially supported by the National Natural Science Foundation of China (No.51906008, No.51822602) and the Fundamental Research Funds for the Central Universities (No. YWF-19-BJ-J-293).